\documentclass[aps,preprint]{revtex4}%
\usepackage{amsfonts}
\usepackage{amsmath}
\usepackage{amssymb}
\usepackage{graphicx}%
\begin{document}
\title{Entropy in the interior of a black hole and thermodynamics}
\author{Baocheng Zhang}
\email{zhangbc.zhang@yahoo.com}
\affiliation{School of Mathematics and Physics, China University of Geosciences, Wuhan
430074, China}

\begin{abstract}
Based on a recent proposal for the volume inside a black hole, we calculate
the entropy associated with this volume and show that such entropy is
proportional to the surface area of the black hole. Together with the
consideration of black hole radiation, we find that the thermodynamics
associated with the entropy is highly possible to be caused by the vacuum
polarization near the horizon.

\end{abstract}
\maketitle

\section{Introduction}

How big is a black hole? This is not an easy question, since the definition of
the volume of the space inside a black hole depends on how the spacetime is
sliced into space and time, unlike the surface area of the black hole that is
the same for all observers. Intuitively, a nice description for the volume
should be slicing invariant, which was made firstly by Parikh \cite{mkp06} and
also discussed by others \cite{dg05,dm09,bl10,cgp11,bl13}. Recently, a
different method was suggested by Christodoulou and Rovelli \cite{cr15} based
on a simple observation that the interior of Schwarzschild black holes is not
static, which leaded to another sensible description for the volume as the
largest volume bounded by the event horizon of a black hole. For a
Schwarzschild black hole, they showed that at late time the volume took such
an expression,
\begin{equation}
V_{CR}\sim3\sqrt{3}\pi M^{2}\upsilon\label{mv}%
\end{equation}
where $M$ is the mass of the black hole and $v$ is the advanced time. In the
paper, we will call the volume as CR volume for brevity. Instantly, this
result was extended to other backgrounds of spacetime \cite{bj15,yco15,yco152}.

Relative to the specific forms for the volume inside a black hole, ones
concern more about its significance, that is, why do we want to investigate
the volume? This was regarded to be relevant to the black hole information
loss paradox in past studies. So it is unavoidable to involve the black hole
radiation \cite{swh74} and the corresponding thermodynamics \cite{jdb73,bch73}
in such studies, but up to now they only were involved with the qualitative
discussions. In this paper, we will try to make some quantitative calculations
to estimate the entropy associated with the CR volume.

As well-known, a black hole can emit the thermal radiation that
makes the black hole have a lifetime $\sim$ $M^{3}$ and thus the CR
volume inside the black hole has an extraordinary form that is
proportional to $M^{5}$. A natural question arises: for such a large
volume, how many field modes can be included in it. Furthermore,
whether these modes are relevant to the interpretation of
Bekenstein-Hawking entropy statistically. When the radiation
happens, the background geometry will be altered according to the
semiclassical Einstein equation, which, in equilibrium, can be
described with the first law of black hole thermodynamics. Thus one
might ask: whether the maximal volume is also included in the
background geometry. If the volume changes, how is the first law
changed? In this paper, we will work on these problems.

The structure of the paper is as follows. Firstly, we will revisit the
definition of the volume inside a black hole by Christodoulou and Rovelli, and
present explicitly the choice of the maximal hypersurfaces \cite{mt80} by
maximal slicing \cite{ewt73,cil11} in the second section. Then, we calculate
the entropy in the volume using the standard statistical method in the third
section. In the forth section, we discuss the first law of black hole
thermodynamics, in particular for the newly obtained entropy, which is related
to the vacuum polarization \cite{fn97} near the horizon of the black hole.
Finally, we will give a conclusion in the five section.

\section{Black hole volume and maximal slicing}

Start with the geometry of a collapsed object as in Ref. \cite{cr15}, which
can be described with the Eddington-Finkelstein coordinates,%
\begin{equation}
ds^{2}=-f(r)dv^{2}+2dvdr+r^{2}d\theta^{2}+r^{2}\sin^{2}\theta d\phi^{2}
\label{ef}%
\end{equation}
where $f(r)=1-\frac{2M}{r}$ and the advanced time $v=t+\int\frac{dr}%
{f(r)}=t+r+2M\ln\left\vert r-2M\right\vert $. In particular, we have taken the
units $G=c=\hbar=k_{B}=1$. The advantage of the coordinates over the static
one is that there is no coordinate singularity at the event horizon. Thus it
can be analytically continued to all $r>0$, which is required for the
description of the geometry of the collapsed matter.

In order to calculate the volume, a proper hypersurface has to be chosen. With
an transformation $v\rightarrow v\left(  T,\lambda\right)  ,r\rightarrow
r\left(  T,\lambda\right)  $, the coordinates (\ref{ef}) becomes $ds^{2}=$
$\left(  -f\left(  \frac{\partial v}{\partial T}\right)  ^{2}+2\frac{\partial
v}{\partial T}\frac{\partial r}{\partial T}\right)  dT^{2}+\left(  -f\left(
\frac{\partial v}{\partial\lambda}\right)  ^{2}+2\frac{\partial v}%
{\partial\lambda}\frac{\partial r}{\partial\lambda}\right)  d\lambda^{2}%
+r^{2}d\theta^{2}+r^{2}\sin^{2}\theta d\phi^{2}$ where we have assumed the
cross term vanishes by taking the transformation properly. In particular, if
the condition $-f\left(  \frac{\partial v}{\partial T}\right)  ^{2}%
+2\frac{\partial v}{\partial T}\frac{\partial r}{\partial T}=-1$ (i.e. this
can be realized by taking $dv=\frac{-1}{\sqrt{-f}}dT+d\lambda,dr=\sqrt{-f}dT$
which also removes the cross term simultaneously) is enforced, one will find
that the hypersurface $\Sigma$: $T=$constant, is that chosen in Ref.
\cite{cr15} where the spherically symmetric hypersurface is taken as the
direct product of a $2$-sphere and an arbitrary curve parameterized by
$\lambda$ in $v$-$r$ plane. In particular, it is noted that the hypersurface
$T=$constant is able to be gotten by $r=$constant according to the
transformation $dr=\sqrt{-f}dT$.

The hypersurface $\Sigma$, as in Ref. \cite{cr15}, is coordinatized by
$\lambda,\theta,\phi$, and the line element of the induced metric on it can be
expressed as%
\begin{equation}
ds_{\Sigma}^{2}=\left(  -f(r)\dot{v}^{2}+2\dot{v}\dot{r}\right)  d\lambda
^{2}+r^{2}d\theta^{2}+r^{2}\sin^{2}\theta d\phi^{2}%
\end{equation}
where the dot represents the partial derivative with regard to the parameter
$\lambda$, and $-f(r)\dot{v}^{2}+2\dot{v}\dot{r}>0$ for the spacelike
hypersurface. The maximal volume can be obtained by the integral,
\begin{equation}
V_{\Sigma}=4\pi\int d\lambda\sqrt{r^{4}\left(  -f(r)\dot{v}^{2}+2\dot{v}%
\dot{r}\right)  }. \label{vi}%
\end{equation}
with a proper choice of the curves. An investigation for geodesics in an
auxiliary manifold gave the maximization condition by choosing the curves
\cite{cr15},
\begin{equation}
r=\frac{3}{2}M, \label{mvc}%
\end{equation}
which, together with Eq. (\ref{vi}), gives the CR volume expressed in Eq.
(\ref{mv}).

On the other hand, the maximization can also be calculated in mathematical
relativity \cite{eg07} where a method called as maximal slicing can lead to
hypersurfaces of maximal volume which have vanishing mean extrinsic curvature,
$K=0$, where $K$ is the trace of the extrinsic curvature of the hypersurface.
Here we show the vanishing $K$ is equivalent to the condition (\ref{mvc}).
According to Ref. \cite{cil11}, we use the coordinates (\ref{ef}) and take the
spacelike hypersurfaces by $r=$constant, since the time and space are regarded
as being interchanged across the horizon of a Schwarzschild black hole. Take
$n$ as the future pointing timelike unit normal to the hypersurfaces
$\Sigma_{r}$,%
\[
n=\sqrt{-f}\left(  \frac{\partial}{\partial r}+\frac{1}{f}\frac{\partial
}{\partial v}\right)  .
\]
It is easy to confirm that $g_{\mu\nu}n^{\mu}n^{\upsilon}=-1$ for the
coordinates (\ref{ef}). It is noted that there is a simple relation between
the divergence of the vector $n$ and the trace of the extrinsic curvature
tensor $K_{\mu\nu}$, $K=-\bigtriangledown\cdot n$ where $\bigtriangledown$ is
the covariant derivative with respect to the spacetime metric. With the given
metric (\ref{ef}), we have%
\[
K=\frac{1}{2\sqrt{-f}}\left(  \frac{\partial f}{\partial r}+4f\frac{1}%
{r}\right)  =0,
\]
which gives the equation $r=\frac{3}{2}M$, consistent with the condition
(\ref{mvc}).

\section{Entropy in the volume}

Since the CR volume is obviously different from the normal volume that might
still exist inside the black hole, it is significant to investigate how many
modes of quantum fields can be included in CR volume. In the paper, we
involves only the scalar field. According to the standard quantum statistical
method \cite{quan}, the number of quantum states in some volume has to be
counted for one certain phase-space which can be labeled here by
$\lambda,\theta,\phi,p_{\lambda},p_{\theta},p_{\phi}$. From the uncertainty
relation of quantum mechanics, $\Delta x_{i}\Delta p_{i}\sim2\pi$, one quantum
state corresponds to a \textquotedblleft cell\textquotedblright\ of volume
$\left(  2\pi\right)  ^{3}$ in the phase-space. Therefore, the number of
quantum states is given by
\begin{equation}
\frac{d\lambda d\theta d\phi dp_{\lambda}dp_{\theta}dp_{\phi}}{\left(
2\pi\hbar\right)  ^{3}}. \label{qs}%
\end{equation}

In order to calculate the integral, we consider a massless scalar field $\Phi$
in the spacetime with the coordinates,
\begin{equation}
ds^{2}=-dT^{2}+\left(  -f(r)\dot{v}^{2}+2\dot{v}\dot{r}\right)  d\lambda
^{2}+r^{2}d\theta^{2}+r^{2}\sin^{2}\theta d\phi^{2}.\label{dcr}%
\end{equation}
It is noted that this metric is equivalent to such an form: $ds^{2}%
=-dT^{2}+H(T)d\lambda^{2}+r\left(  T\right)  ^{2}d\theta^{2}+r\left(
T\right)  ^{2}\sin^{2}\theta d\phi^{2}$, which means that it is not static for
the defined time $T$ in the interior of the black hole. For this reason,
sometimes the interior of the black hole is interpreted as a cosmological
model (see also Ref. \cite{cjr09}), and it evolves towards the singularity of
the black hole. It is not entirely clear what should be done for the
statistics in a dynamic background, but fortunately, our calculation is made
at $v>>M$ and $r=\frac{3}{2}M$. For the maximal slicing, the slices accumulate
on a limiting hypersurface $r=\frac{3}{2}M$ when $t$ is large enough (that is
guaranteed by $v>>M$) \cite{ewt73}. That is to say, near the maximal
hypersurface, the proper time between two neighbouring hypersurfaces tends to
zero as $t$ increases, so nearly no evolution happened there. Thus our
statistical calculation is not affected by the non-static character of the
metric, since it is calculated on approximately $T=$constant which is the
hypersurface that leads to the CR volume. Therefore, in what follows we will
use the common method in the curved spacetime to discuss the motion of scalar
field in the interior of the black hole.

Using the WKB approximation, the field $\Phi$ can be written as $\Phi
=\exp[-iET]\exp[iI\left(  \lambda,\theta,\phi\right)  ]$, and then
substituting it into the Klein-Gordon equation in curved spacetime, $\frac
{1}{\sqrt{-g}}\partial_{\mu}\left(  \sqrt{-g}g^{\mu\nu}\partial_{\nu}%
\Phi\right)  =0$, we obtain%
\begin{equation}
E^{2}-\frac{1}{-f(r)\dot{v}^{2}+2\dot{v}\dot{r}}p_{\lambda}^{2}-\frac{1}%
{r^{2}}p_{\theta}^{2}-\frac{1}{r^{2}\sin^{2}\theta}p_{\phi}^{2}=0,
\end{equation}
where $p_{\lambda}=\frac{\partial I}{\partial\lambda},p_{\theta}=$
$\frac{\partial I}{\partial\theta},p_{\phi}=\frac{\partial I}{\partial\phi}$.
Thus, following the Eq. (\ref{qs}), the number of quantum states with the
energy less than $E$ is obtained as%
\begin{align}
g\left(  E\right)   &  =\frac{1}{\left(  2\pi\right)  ^{3}}\int d\lambda
d\theta d\phi dp_{\lambda}dp_{\theta}dp_{\phi}\nonumber\\
&  =\frac{1}{\left(  2\pi\right)  ^{3}}\int\sqrt{-f(r)\dot{v}^{2}+2\dot{v}%
\dot{r}}d\lambda d\theta d\phi\int\sqrt{E^{2}-\frac{1}{r^{2}}p_{\theta}%
^{2}-\frac{1}{r^{2}\sin^{2}\theta}p_{\phi}^{2}}dp_{\theta}dp_{\phi}\nonumber\\
&  =\frac{1}{\left(  2\pi\right)  ^{3}}\int\sqrt{-f(r)\dot{v}^{2}+2\dot{v}%
\dot{r}}d\lambda d\theta d\phi\left(  \frac{2\pi}{3}E^{3}r^{2}\sin
\theta\right) \nonumber\\
&  =\frac{E^{3}}{12\pi^{2}}\left[  4\pi\int d\lambda\sqrt{r^{4}\left(
-f(r)\dot{v}^{2}+2\dot{v}\dot{r}\right)  }\right] \nonumber\\
&  =\frac{E^{3}}{12\pi^{2}}V_{CR},
\end{align}
where the relation $p_{\lambda}=\sqrt{-f(r)\dot{v}^{2}+2\dot{v}\dot{r}}%
\sqrt{E^{2}-\frac{1}{r^{2}}p_{\theta}^{2}-\frac{1}{r^{2}\sin^{2}\theta}%
p_{\phi}^{2}}$ is used in the second line, the integral formula $\int\int
\sqrt{1-\frac{x^{2}}{a^{2}}-\frac{y^{2}}{b^{2}}}dxdy=\frac{2\pi}{3}ab$ is used
in the third line, and in the final line we have used the condition
(\ref{mvc}). As expected, the number of quantum states is proportional to the
CR volume, but this is still different from a normal situation, because the CR
volume is the result of the curved spacetime.

Temporarily ignoring the exotic feature of CR volume, we can continue to
calculate the free energy at some inverse temperature $\beta$,
\begin{align}
F\left(  \beta\right)   &  =\frac{1}{\beta}\int dg\left(  E\right)  \ln\left(
1-e^{-\beta E}\right) \nonumber\\
&  =-\int\frac{g\left(  E\right)  dE}{e^{\beta E}-1}\nonumber\\
&  =-\frac{V_{CR}}{12\pi^{2}}\int\frac{E^{3}dE}{e^{\beta E}-1}\nonumber\\
&  =-\frac{\pi^{2}V_{CR}}{180\beta^{4}}.
\end{align}

Furthermore, the entropy is obtained as%
\begin{equation}
S_{CR}=\beta^{2}\frac{\partial F}{\partial\beta}=\frac{\pi^{2}V_{CR}}%
{45\beta^{3}}, \label{cre}%
\end{equation}
which looks like the entropy in the normal volume.

Now we turn to the CR volume again, but under the consideration with Hawking
radiation. Since Hawking radiation is thermal, the loss mass rate of a
Schwarzschild black hole can be given by the Stefan--Boltzmann law,%
\begin{equation}
\frac{dM}{dv}=-\frac{1}{\gamma M^{2}},\gamma>0, \label{sbl}%
\end{equation}
where $\gamma$ is a constant whose value does not influence the discussion in
the paper. In Ref. \cite{yco152}, it is discussed that the large volume is
remained until the final stage of black hole evaporation. Here we are
concerned about the time that the radiation can last. Thus, for a black hole
with the mass $M$, we have%
\begin{equation}
v\sim\gamma M^{3},
\end{equation}
which also satisfies the requirement in Ref. \cite{cr15}, $v>>M$. Then using
the Eqs. (\ref{mv}) and (\ref{cre}), and considering the temperature for a
Schwarzschild black hole, $\beta=T^{-1}=8\pi M$, one find the entropy
\begin{equation}
S_{CR}\sim\frac{\left(  3\sqrt{3}\gamma\right)  M^{2}}{\left(  45\times
8^{3}\right)  }=\frac{\left(  3\sqrt{3}\gamma\right)  }{\left(  90\times
8^{4}\right)  \pi}A,
\end{equation}
where $A=16\pi M^{2}$ is the surface area of the Schwarzschild black hole.
This is a surprising and intriguing result that the entropy from quantum
theory in the CR volume is proportional to the surface area of the black hole
horizon that bounds the volume. Note that the result is dependent on the
validity of the relation (\ref{sbl}), which was shown \cite{sm95} to hold so
long as the mass of a Schwarzschild black hole is greater than the Planck
mass. Moreover, the exact description of final stage of black hole evaporation
has not existed, but some specific considerations such as General Uncertainty
Relation (i.e. see the review \cite{sh13}) implied that at the final stage,
the loss mass rate becomes $\frac{dM}{dv}\sim M$ \cite{lx02}, which is
inconsistent with the requirement for the calculation of the CR volume, beyond
our problem obviously. On the other hand, although the entropy $S_{CR}$ is
proportional to the surface area, a rough estimation finds that the parameter
before the area is much smaller than $\frac{1}{4}$. Then a natural problem:
what is the entropy $S_{CR}$? Whether it will be included in the first law of
black hole thermodynamics. In what follows, we will discuss it.

\section{First law and vacuum polarization}

Since the derivation of CR volume is in the case of $v>>M$, which means the
black hole has formed by the collapse of the matter and is static for the
external observers. Thus, due to the Hawking radiation, the first law of black
hole thermodynamics reads as%
\begin{equation}
dM=TdS_{H},
\end{equation}
where the entropy $S_{H}=\frac{1}{4}A$. This relation was obtained
\cite{swh74,jdb73} by the analogy of black hole physics to thermodynamics
which will reach an equilibrium state after the relaxation processes are
completed. The change of the entropy $S_{H}$ is equivalent to the change of
the surface area, so the entropy $S_{CR}$ in the CR volume will be changed.
Although our earlier analysis shows that the entropy $S_{CR}$ is proportional
to the surface area, its original form in Eq. (\ref{cre}) is closely related
to the CR volume. So in equilibrium, the thermal process requires a term
relevant to the change of CR volume. According to the general thermodynamics,
such term should be written as $PdV_{CR}$ where $P$ is the pressure. Thus, we
expect that the first law can be written as
\begin{equation}
dM=TdS-PdV_{CR},
\end{equation}
where $S=S_{H}+S_{CR}$. In order to remain the validity of the first law, a
relation is required, $TdS_{CR}\sim PdV_{CR}$. Now, the vital problem is that
from where and how much the pressure is, after the collapsed matter had been
concentrated into the singularity (it is also noted that some studies pointed
out that the nonlocal effect will prevent the black hole from collapsing into
a physical singularity, i.e. see Ref. \cite{ss14}).

In the original calculation of Hawking \cite{swh74}, the concept of particles
are used in the asymptotically flat region far from the black hole where they
can be unambiguously defined. While the particle flux carries away the
positive energy, an accompanying flux of negative energy goes into the hole
across the horizon, which can only be understood by the zero point
fluctuations of local energy density in the quantum theory. This phenomena,
called also as vacuum polarization, will play an important role in the
neighborhood of the black hole.

The vacuum polarization is usually considered in the semi-classical Einstein
gravity, in which the fluctuations of gravitational field are small and so is
the expectation value of the energy-momentum tensor of the relevant quantized
fields in the chosen vacuum state. By solving the modified Einstein equation,
$G_{\mu\nu}=8\pi\left\langle T_{\mu\nu}\right\rangle $, the quantum pressure
at the horizon caused by the vacuum polarization is given by
\cite{wgu76,pc80,dnp82,te83},
\begin{equation}
P=\frac{1}{\left(  90\times8^{4}\right)  \pi^{2}}\frac{1}{M^{4}}.
\end{equation}
It is noted that the CR volume is calculated for the hypersurface $r=\frac
{3}{2}M$, while $P$ is at the horizon. But it is not difficult to clarify
this, since the CR volume is just the volume for the black hole, and the
boundary for the volume is still the horizon. This point was also seen in Ref.
\cite{cr15}, by the parameter $\lambda$ that take the range covering the
region from $r=0$ to $r=2M$. In particular, some recent interesting studies
considered the pressure as stemming from the cosmological constant, i.e. see
Refs. \cite{krt09,bpd11,km14}, but they are different from ours since the
cosmological constant is not involved here.

The term $PdV_{CR}$ is calculated as%
\begin{equation}
PdV_{CR}=\frac{5\left(  3\sqrt{3}\gamma\right)  }{\left(  90\times
8^{4}\right)  \pi}dM,
\end{equation}
which is on the level of $10^{-5}$ for the parameter before $dM$, and is very
small, as expected. The term $TdS_{CR}$ is also easy to be estimated as%
\begin{equation}
TdS_{CR}=\frac{4\left(  3\sqrt{3}\gamma\right)  }{\left(  90\times
8^{4}\right)  \pi}dM\sim10^{-5}dM.
\end{equation}
It is seen that the term $TdS_{CR}$ is not equal exactly to the term
$PdV_{CR}$, but fundamentally they can be canceled on the level of $10^{-5}$.
The reason is that the values of the pressure $P$ and the temperature $T$ is
not so exact in the present consideration. Actually, a direct evidence that
the thermodynamics in the CR volume is caused by vacuum polarization is that
the term $PdV_{CR}$ is not dependent on the black hole parameter $M$, as
obtained for the term $TdS_{CR}$.

Finally, it is noted that if taking such a term $V_{CR}dP$, as made
in Refs. \cite{krt09,bpd11,km14}, it can balance exactly the change
of thermodynamics caused by $TdS_{CR}$, but no evidence shows that
why the pressure here should be changed, so the exact cancelation
might be occasional. In fact, vacuum polarization causes the quantum
correction to the usual black hole thermodynamics, so the
corresponding thermodynamic variables would be corrected, which is
the reason why the terms $PdV_{CR}$ and $TdS_{CR}$ cancel each other
out approximately.

\section{Conclusion}

In the paper, we have investigated the relation between the derivation of the
volume of the space inside a Schwarzschild black hole defined by Christodoulou
and Rovelli and the maximal slicing, and found explicitly that the CR volume
was just obtained for the hypersurface whose mean extrinsic curvature is zero.
We have also calculated the entropy in the CR volume through counting the
number of quantum states in the volume with a standard statistical method.
Differently from the normal situation, the entropy associated with the CR
volume is proportional to the surface area of the black hole, but the
parameter is much smaller than that required for the Bekenstein-Hawking
entropy. The small parameter has also been interpreted in the paper, from the
perspective of black hole thermodynamics, for which a suggestive result is
given that the thermodynamics associated with the entropy in the CR volume is
caused by the vacuum polarization near the horizon, since the matter has
collapsed into the singularity when we investigate this phenomena. Thus, our
result verifies further the relation between black hole physics and quantum
theory again.

\section{Acknowledge}

The author would like to thank Prof. C. Rovelli for reading this paper and his
positive comments. This work is supported by Grant No. 11374330 of the
National Natural Science Foundation of China and by Open Research Fund Program
of the State Key Laboratory of Low-Dimensional Quantum Physics and by the
Fundamental Research Funds for the Central Universities, China University of
Geosciences (Wuhan)(CUG150630).

\end{document}